\documentclass[11pt]{article}
\usepackage{moriond}

\def\Journal#1#2#3#4{{#1} {\bf #2}, #3 (#4)}

\def\PLB{{\em Phys. Lett.}  B}
\def\PRL{\em Phys. Rev. Lett.}
\def\PRD{{\em Phys. Rev.} D}

\def\be{\begin{equation}}
\def\ee{\end{equation}}
\def\bea{\begin{eqnarray}}
\def\eea{\end{eqnarray}}

\newcommand{\neb}{$\bar{\nu}_{\rm{e}} \; $}

\newcommand{\Photo}{}

\begin{document}
\vspace*{4cm}

\title{DOUBLE CHOOZ: RESULTS TOWARDS THE NEAR DETECTOR PHASE}

\author{ P. NOVELLA }

\address{APC-CNRS,\\
10, rue Alice Domon et Léonie Duquet, 75205 Paris, France}

\maketitle

\abstracts{Since the first indication in 2011 of a non-vanishing value of $\theta_{13}$ using reactor neutrinos by Double Chooz, this collaboration has developed new analyses leading to an increased precision on $\theta_{13}$ and different cross-checks of the oscillation results. Beyond the update of the analysis based on neutron captures on Gd in 2012, Double Chooz has also released a consistent measurement of $\theta_{13}$ by means of neutron captures on H. The combination of the n-Gd and n-H data samples, as well as 7.23 days of reactor off data, in a global rate+shape analysis yields $\sin^22\theta_{13} = 0.109\pm 0.035$. This measurement has been validated with a new background-model-independent approach, which compares the observed and expected neutrino rates as a function of reactor power. This Reactor Rate Modulation analysis yields $\sin^22\theta_{13} = 0.102\pm 0.043$ when combining the n-Gd and n-H samples. The Double Chooz collaboration is currently developing a new analysis increasing the signal-to-background ratio while reducing the background and detection systematics. This new approach is being prepared as a first step of the new phase of the experiment that will begin in summer 2014 with the operation of a near detector.}

\section{Oscillation analyses in Double Chooz}

The Double Chooz reactor neutrino experiment aims at measuring the neutrino oscillation in the interference sector with a precision of 10\% on $\sin^22\theta_{13}$. This parameter is determined by means of the observation of a deficit in the electron antineutrino (\neb) flux at a distance of about 1 km from the two Chooz reactors (8.54 GW$_{th}$) in France. The so-called far detector is placed in a laboratroy close to the maximal oscillation distance and providing enough shielding (300 m.w.e.) against cosmic rays. A second identical detector (near detector)  is being built in a new experimetal site (115 m.w.e), located about 400 m away from the reactor cores, and will start operation by summer 2014. 


The Double Chooz detectors design is optimized to reduce backgrounds. The detectors consist of a set of concentric cylinders and an outer plastic scintillator muon veto (outer veto or OV) on the top. The innermost volume (target or TG) contains about 10 tons of Gd-loaded (0.1\%) liquid scintillator inside a transparent acrylic vessel. This volume is surrounded by another acrylic vessel filled with unloaded scintillator (the  gamma-catcher or GC). The GC is in turn contained within a third volume (buffer tank or BT) made of stainless steel and filled with mineral oil. The walls of BT are equipped with an array of 390 10'' photomultiplier tubes (PMTs). Finally, the outer volume containing the TG, GC and BF (inner detector) is a stainless steel vessel covered with 78 8'' PMTs and filled with scintillator, playing the role of an inner muon veto. The \neb are detected via the inverse beta decay (IBD) reaction: ${\bar \nu}_e p \to e^+ n$. An IBD event is identified by the correlated (in both time and space) prompt-$e^+$ and delayed-$n$-capture energy depositions. The \neb energy is accurately measured using the positron energy deposition. The neutrons generated in the IBD processes are captured mainly in Gd (when taking place in the TG) and H (either in the TG or the GC). The energy despositions upon the n-captures are $\sim$8 MeV and $\sim$2.2 MeV for Gd and H, respectively.


In reactor neutrino experiments like Double Chooz, a comparison between the observed and expected antineutrinos, in terms of both rate and energy spectrum, provides a clean measurement of $\theta_{13}$. In 2011, Double Chooz published a first indication of a non-vanishing value of $\theta_{13}$ with reactor neutrinos \cite{dcngdI}, using n-captures in Gd and exploiting both the rate and spectral shape information. The oscillation results were updated and improved ($\sim3\sigma$ measurement) in 2012 \cite{dcngdII}, and confirmed with an independent \neb statistical sample obtained with n-captures in H \cite{dcnhII}. While this first analyses depend on a background model, a new background-model-independent $\theta_{13}$ measurement has been released in 2013 \cite{dcrrmII}. Finally, the statistical samples obtained with Gd and H n-captures have been combined in both analysis approaches in order to boost the precision on $\theta_{13}$. A description of all these analyses and results follows.

\section{Detecting \neb in the Double Chooz far detector}


The \neb candidate events are selected for neutron captures on Gd (nGd) and H (nH). The Gd analysis \cite{dcngdII} has the advantage of the high energy delayed event, above the region of natural radioactivity, and the short coincidence time between the prompt and delayed signals ($\Delta$T$\sim30\mu$s). These features lead to a large signal to background ratio S/B ($\sim$17). On the other hand, the H analysis provides a independent statistical sample, although with larger background (S/B$\sim$1) and efficiency systematics. Since the nH approach accounts for the events within the GC, the statistical sample is about a factor three larger than the nGd one. The \neb candidate selection has been described in \cite{dcngdII,dcnhII}, being most of the cuts common in the nGd and nH analyses. First, in order to reduce the background caused by instrumental light production, it is required that the light signal arrives uniformly in space and simultaneously in time at the PMTs. In addition, the triggers collected within 1 ms following a tagged muon are also rejected to reduce the muon-induced backgrounds. The search of \neb is then applied over the remaining events. The prompt energy window is 0.7-12.2 MeV in both analyses, while the delayed energy window is 6-12 MeV for the nGd selection and 1.5-3 MeV for the nH one. The $\Delta$T cuts are 2-100 $\mu$s and 10-600 $\mu$s, for the nGd and nH samples, respectively. Any extra trigger around the coincident signals is allowed. In both selections, candidates with a prompt signal coincident in time with an OV trigger are rejected. To reduce the number of accidental coincidences, much larger in the H analysis due to the lower delayed energy window, a spatial cut is applied in the H selection: the prompt and delayed vertexes are required to be within a maximum distance of 0.9 m. To reduce the correlated background in the nGd analysis (the dominant one), candidates within a 0.5 s window after a high energy muon (E$>$600 MeV) crossing the inner detector are also rejected. This leads to a shorter live time in the nGd analysis with respect to the nH approach.

\begin{figure}
\begin{center}
\includegraphics[scale=0.8]{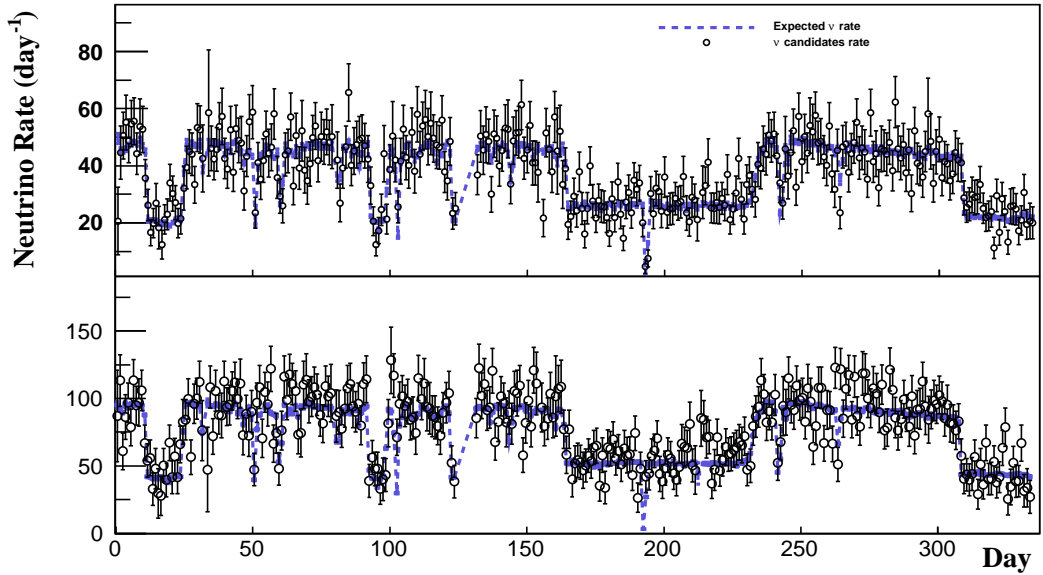}
\caption{ Observed \neb candidate rate (black points) as a function of time, superimposed to the expected rate of \neb events in absence of oscillation. Top: \neb candidates (backgrounds not subtracted) according to the nGd selection. Bottom:  \neb candidates (accidental background subtracted) according to the nH selection.
\label{fig:cands}}
\end{center}
\end{figure}


There are three main backgrounds in the Double Chooz experiment: 1) random coincidences (accidental background), 2) correlated signals produced by fast neutrons (fast-n) and stopping muons (stop-$\mu$), and 3) $\beta-n$ emitters, like $^{9}$Li or like $^{8}$He, produced by spallation muons crossing the inner part of the detector (cosmogenics). With the exception of the energy spectrum of the cosmogenic events, both the rate and the energy of the different background sources are derived from the data themselves, according to the analyses described in \cite{dcngdII,dcnhII}. The accidental, fast-n/stop-$\mu$ and cosmogenic background rates are computed to be 0.26$\pm$0.02, 0.67$\pm$0.20 and 1.25$\pm$0.54 event/day, respectively, for the nGd selection and 73.45$\pm$0.16, 2.50$\pm$0.47 and 2.84$\pm$1.15 event/day for the nH selection. Among other $\theta_{13}$ reactor neutrino experiments, Double Chooz is unique as data can be collected with all reactors off. The analysis of the reactor-off data taken during a total live time of 7.23 days provides a direct measurement of the total background rate. Applying the selection used in the oscillation analysis, 7 (599) neutrino candidates are found within the reactor-off period in the nGd (nH) selection. This number accounts for both background events and residual neutrinos generated when the reactors have been brought down.  In order to evaluate the residual neutrino spectrum in the reactor-off period, a dedicated simulation of the reactor cores evolution has been performed. The total number of expected detected neutrino is 1.4$\pm$0.42 (3.7$\pm$1.1) in the nGd (nH) sample. By subtracting the residual neutrinos to the observed events, a total background rate of 1.0$\pm$0.4 (11.3$\pm$3.4) events/day is derived for the nGd (nH) selection. This is consistent within  2$\sigma$ with the background estimates derived from reactor-on data (2.0$\pm$0.6 and 5.8$\pm$1.3 event/day for nGd and nH, respectively), thus validating the background model used in the oscillation analysis. 


In the current analyses, the data samples in \cite{dcngdII,dcnhII} are used along with the extra reactor-off sample collected in 2012 \cite{dcoffoff}, which increases the total reactor-off run time to 7.53 days. Within the corresponding total live time of 233.93 days (246.4 days), 8257 (36883) candidates (including accidental background) were found according to the n-Gd (n-H) selection, 8 (599) of which were observed during the reactor-off period. Fig. \ref{fig:cands} shows the observed rate of \neb candidates, superimposed to the expected rate of events in absence of oscillation. A summary of the uncertainties on the expected number of events in shown in Tab.~\ref{tab:sys}, along with the the statistical error of the nGd and nH samples. The dominant systematic uncertainty is the reactor-related one (1.8\%), which will be almost fully canceled in the second phase of the experiment using the near detector.

\begin{table}[t]
\caption{Summary of the normalization uncertainties for the Double Chooz Gd and H-analysis}
\label{tab:sys}
\vspace{0.4cm}
\begin{center}
\begin{tabular}{|c|c|c|}
\hline
Source       & nGd selection & nH selection\\ \hline
Reactor flux & 1.8\%         & 1.8\%  \\ 
Efficiency   & 1.0\%         & 1.6\%  \\ 
Background  & 1.6\%         & 1.7\%  \\ \hline
Statistics   & 1.2\%         & 1.08\%  \\ \hline
\end{tabular}
\end{center}
\end{table}

\section{Rate+Shape oscillation results with n-Gd and n-H samples}

\begin{figure}
\begin{center}
\includegraphics[scale=0.39]{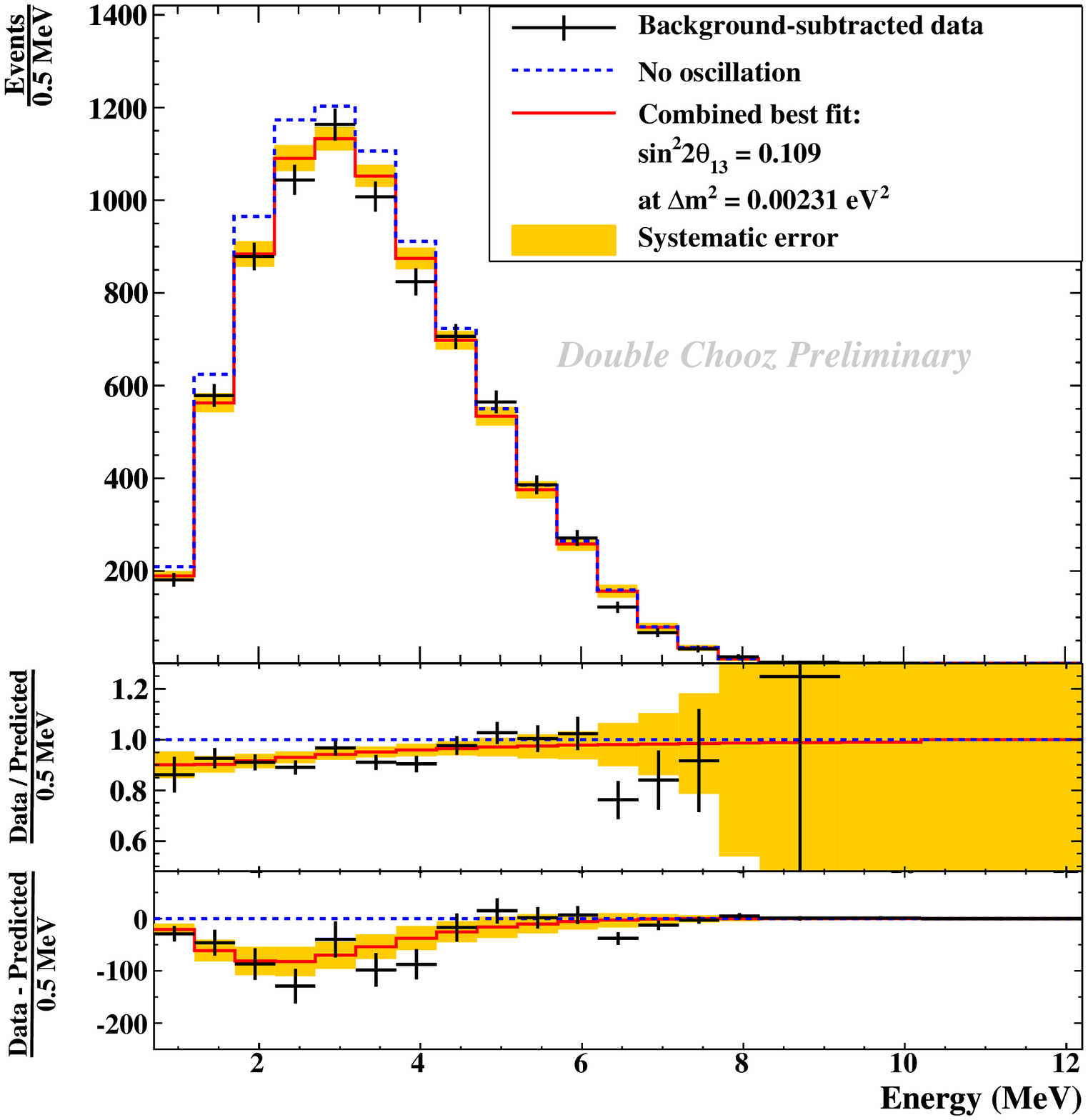}
\includegraphics[scale=0.39]{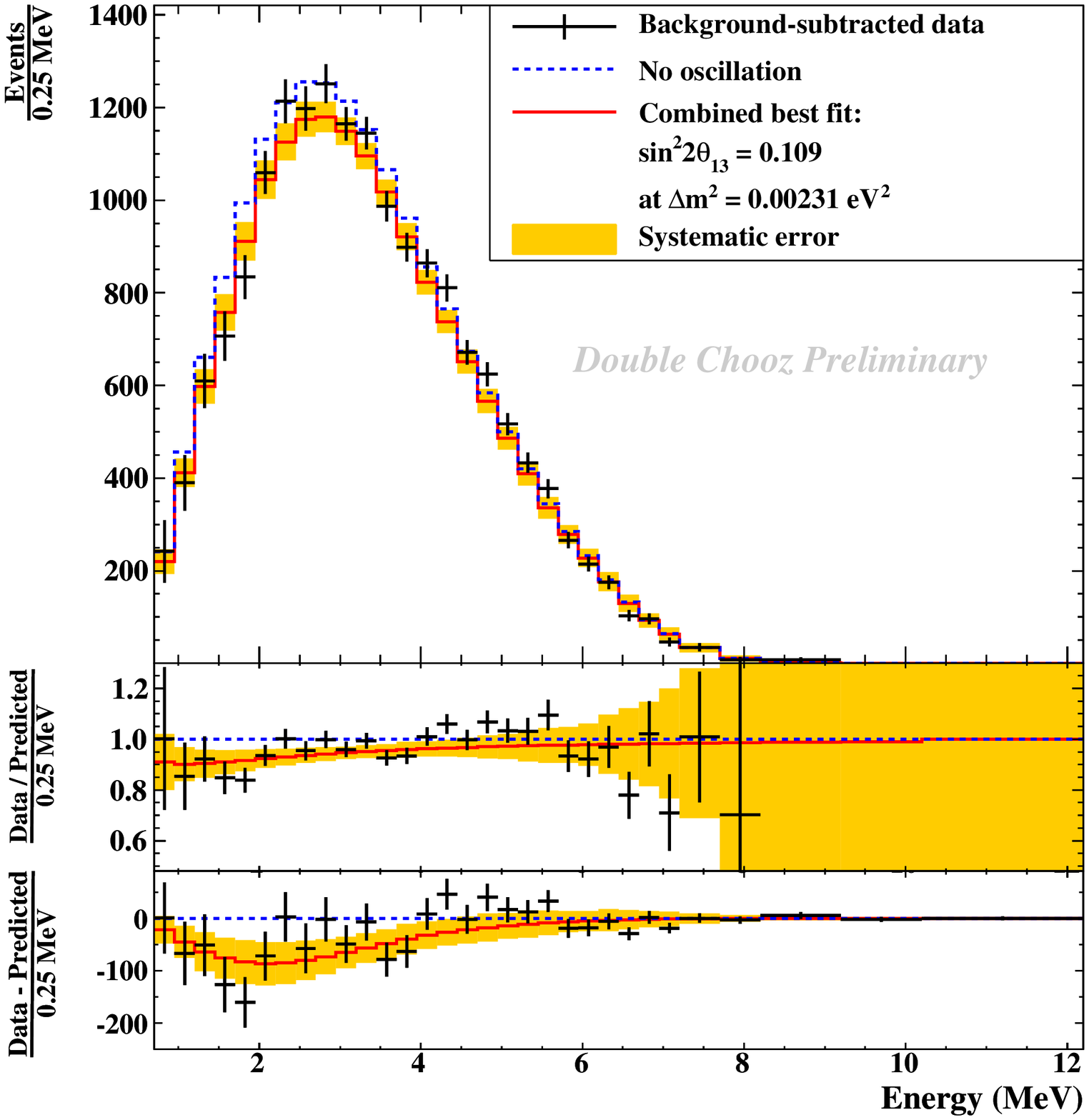}
\caption{ Rate+shape combined fit using n-Gd (right) and n-H (left) \neb candidates. Top: background-subtracted data (black points with statistical error bars) superimposed on no-oscillation prediction (dashed blue line) and combined best fit (red line). Backgrounds are subtracted at their best-fit rates. Gold bands indicate systematic errors in each bin. Bottom: data minus no-oscillation prediction (black points with statistical error bars) superimposed on no-oscillation prediction (dashed blue line) and combined best fit (solid red line). Gold bands indicate systematic errors in each bin.
\label{fig:RScomb}}
\end{center}
\end{figure}

The \neb candidate samples with neutron captures on Gd and H, which are statistically independent, are analyzed separately with the selections and livetime described above. The oscillation analysis is based on a $\chi^2$ fit to the observed \neb rate and energy spectrum, which are compared to the expectation based on Monte-Carlo simulations, described in \cite{dcngdII}, and the background model. In the nGd (nH) oscillation analysis, the data are separated into 18 (31) variably sized bins between 0.7 and 12.2 MeV. A prediction of the observed number of signal and background events is constructed for each energy bin. The associated systematics and statistical uncertainties are propagated to the fit through a covariance matrix and pull terms in the $\chi^2$ function. While the covariance matrix accounts for the statistical uncertainty and the energy spectrum and detection efficiency systematics, the cosmogenic and fast-n/stop-$\mu$ normalization and the energy scale uncertainties are introduced as pulls. Thanks to the energy spectrum (in particular between 8 and 12 MeV where the number of predicted neutrino events is very small), some information about the background rates can be extracted from the fit. This is why the central values of the cosmogenic and fast-n/stop-$\mu$ backgrounds are allowed to vary during the minimization procedure by means of nuisance multiplicative factors (pulls). The same principle applies to the energy scale error, which is also allowed to vary linearly constrained by its uncertainty (1.3\%). For the mass splitting, $\Delta m^2_{31}$ the value from the MINOS measurement of (2.32$\pm$0.12)$\times 10^{−3}$ eV$^2$ \cite{minos} is used. 

According to the above rate+shape analysis, consistent values for $\sin^22\theta_{13}$ are found in \cite{dcngdII} and \cite{dcnhII} using the nGd and nH candidates samples, respectively. The best fit values yield $\sin^22\theta_{13} = 0.109\pm0.039$ for the nGd analysis and $\sin^22\theta_{13}= 0.097\pm0.048$ for the nH analysis. Although the later result is significantly worse due to the larger systematics, both samples can be combined in order to increase the precision on $\theta_{13}$. A preliminary nGd+nH combined fit has been performed, including the correlation of systematic errors and the background constraints as coming out of the reactor-off measurement.  A best fit value of $\sin^22\theta_{13}$ = 0.109 $\pm$ 0.035 is obtained with the rate+shape fit, while $\sin^22\theta_{13} = 0.107 \pm 0.045$ is derived from a fit accounting only for the rate information. Fig.~\ref{fig:RScomb} shows the results of the rate+shape nGd+nH combined fit.

\section{RRM analysis: a background model independent approach}

\begin{figure}
\begin{center}
\includegraphics[scale=0.39]{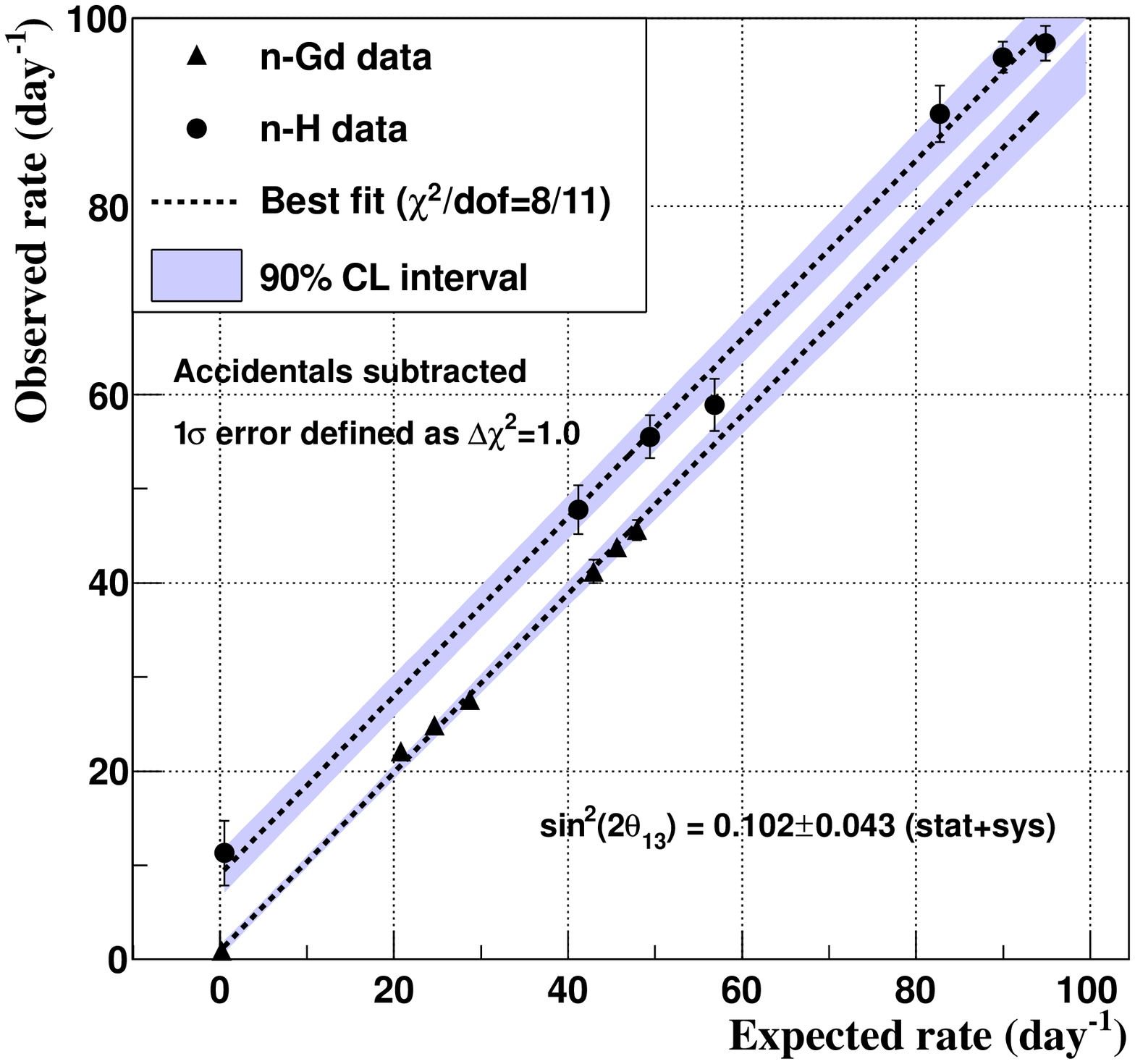}
\includegraphics[scale=0.39]{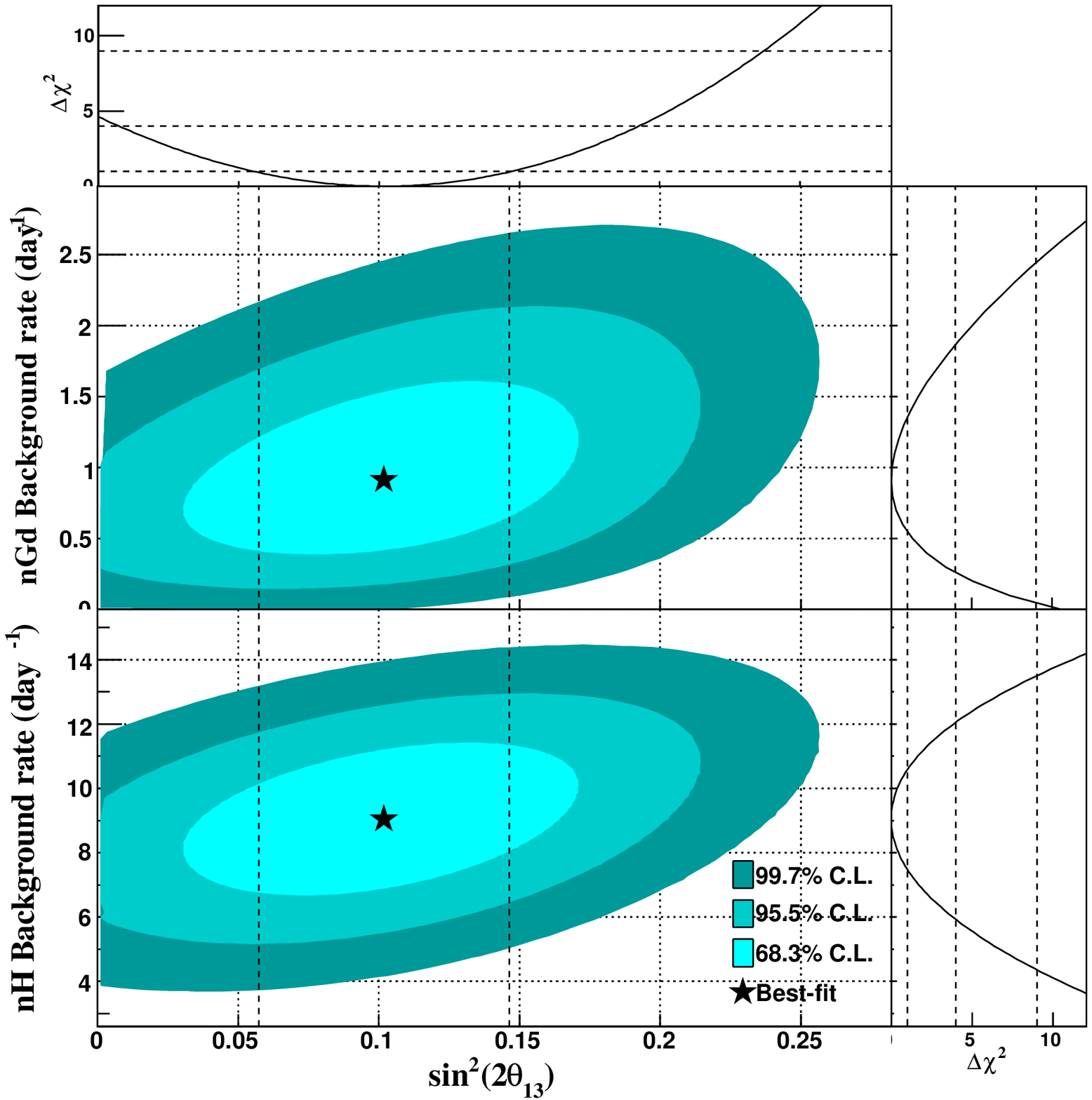}
\caption{ RRM combined fit using n-Gd and n-H \neb candidates. Right: observed \neb candidate rate versus the unoscillated expected rate, with fit results superimposed. Left: best fit value and confidence regions for $\sin^22\theta_{13}$ and the total background rate in the nGd and nH samples.
\label{fig:RRM}}
\end{center}
\end{figure}

While the above oscillation results rely on a background model derived from reactor-on data, the reactor rate modulation (RRM) analysis is a background model independent approach \cite{dcrrmII}. Both $\theta_{13}$ and the total background rate are derived without model assumptions on the background by a global fit to the observed antineutrino rate as a function of the non-oscillated expected rate for different reactor power conditions. Although the RRM fit with only reactor-on data does not achieve a competitive precision on $\theta_{13}$, it provides an independent determination of the total background rate (2.8$\pm$2.0 event/day in a 2-parameter fit). This rate is consistent with the Double Chooz background model and with the measurement of the total background from the 7.53 days of reactor-off data \cite{dcoffoff}. As this sample provides the most precise determination of the total background rate in a model independent way, it is introduced in the RRM analysis in order to improve the results on $\theta_{13}$, which remains as the only free parameter in the fit. The best fit value of $\sin^{2}2\theta_{13}$=0.107$\pm$0.049 is found by analyzing the nGd \neb candidates, while $\sin^{2}2\theta_{13}$=0.091$\pm$0.078 is obtained with the nH sample. Finally, the precision on $\theta_{13}$ is further improved by combining the nGd and nH \neb samples:  $\sin^22\theta_{13}=0.102\pm$0.028(stat.)$\pm$0.033(syst.). The outcome of the RRM fit is consistent within 1$\sigma$ with the other published results for $\theta_{13}$, yielding a competitive precision. Beyond the cross-check of the background estimates in the Double Chooz oscillation analyses, the RRM analysis provides, for the first time, a background model independent determination of the $\theta_{13}$ mixing angle. The combined nGd+nH RRM fit is shown in Fig.~\ref{fig:RRM}.

\section{A new analysis for the Near Detector Phase}

During the first phase of the Double Chooz experiment, operating only the far detector, the precision on $\theta_{13}$ is limited by the reactor-related uncertainties. In preparation for the second phase, starting in summer 2014 with the operation of the near detector, the work of the collaboration is currently focused on the reduction of all other systematics as the reactor ones will be almost fully canceled. The new analysis being developed cosists of two main building blocks: 1) the improvement of the energy reconstruction by means of new calibration techniques, and 2) the definition of a completely new \neb selection procedure boosting the S/B ratio. Both contributions will lead to the reduction of the systematic uncertainties associated to the backgrounds, the detection efficiency and the energy scale. After three years of data taking with both the far and near detectors, a final precision on $\theta_{13}$ of 10\% is expected.

\section{Conclusions}

In summary, Double Chooz has released four different $\theta_{13}$ results by means of two statistical samples (nGd and nH) and two analysis approaches (rate+shape fit with background inputs and RRM). These results are correlated only through the common detection and reactor-related systematics,  and have being obtained as follows: 1) with n-Gd candidates in \cite{dcngdII}, 2) with n-H candidates in \cite{dcnhII}, 3) with n-Gd candidates and the RRM analysis \cite{dcrrmII}, and 4) with n-H candidates and the RRM analysis \cite{dcrrmII}. The remarkable agreement of the 4 measurements, each having different systematics, demonstrates the robustness of the results and the high precision background knowledge. In order to boost the precision, the nH and nGd measurements have been combined, considering all correlated terms. A nGd+nH combined rate+shape fit (background-model dependent) yields $\sin^22\theta_{13}$ = 0.109 $\pm$ 0.035, while a nGd+nH RRM fit (background-model independent) obtains $\sin^22\theta_{13}$ = 0.102 $\pm$ 0.043.

The second phase of the experiment will start in summer 2014, when the near detector will be operative. Thanks to the cancellation of the reactor-related systematics provided by the near detector data and to the new analysis being currently developed, a final precision of 10\% on $\sin^2 2\theta_{13}$ is expected to be reached after 3 years of data taking.

\end{document}